# Impact of Economic Uncertainty, Geopolitical Risk, Pandemic, Financial & Macroeconomic Factors on Crude Oil Returns: An Empirical Investigation

Sarit Maitra
Alliance Business School
Alliance University, Bengaluru, India
sarit.maitra@gmail.com

**Abstract**

This study aims to use simultaneous quantile regression (SQR) to examine the impact of macroeconomic and financial uncertainty including global pandemic, geopolitical risk on the futures returns of crude oil (ROC). The data for this study is sourced from the FRED (Federal Reserve Economic Database) economic dataset; the importance of the factors have been validated by using variation inflation factor (VIF) and principal component analysis (PCA). To fully understand the combined effect of these factors on WTI, study includes interaction terms in the multi-factor model. Empirical results suggest that changes in ROC can have varying impacts depending on the specific period and market conditions. The results can be used for informed investment decisions and to construct portfolios that are well-balanced in terms of risk and return. Structural breaks, such as changes in global economic conditions or shifts in demand for crude oil, can cause return on crude oil to be sensitive to changes in different time periods. The unique aspect ness of this study also lies in its inclusion of explanatory factors related to the pandemic, geopolitical risk, and inflation.

1. **Introduction**

Crude oil is the most traded commodity globally, with West Texas Intermediate (WTI)[1] and North Sea Brent crude (Brent) being the most widely used benchmarks. The global oil market is valued over $1.7 trillion, making it crucial for creating an ideal portfolio (Nasir et al., 2018; Sarwar et. al., 2019). However, uncertainties like supply and demand, geopolitical events, and economic conditions affect the Return in Crude Oil (ROC) from an investment perspective. There is evidence to believe that efficient market hypothesis has been the failure in most of the energy market (Liu & Lee, 2018). The price of crude oil directly or indirectly impacts all aspects of the economy, and energy market shocks can disrupt the economy and financial system.

The relationship between crude oil price (COP) and macroeconomic variables is influenced by factors such as global economic policy uncertainty (GEPU), geopolitical risk (GPR), global pandemic effect (WUPI), and global price of energy index (GPE). COP remains a significant determinant of economic factors, including inflation rates, which affect the global economy (Tiwari et. al., 2019). The impact of WUPI on COP can be influenced by various variables, such as consumption habits and global supply system disruptions. Understanding the relationship between ROC, excess market return, volatility index, inflation, WUPI, and GPR is crucial for developing hedging methods (Khan et al., 2017; Ferrer et al., 2018). The distribution of ROC can be heterogeneous due to distinct risk-return profiles of oil investments. However, there is limited research on the asymmetric effects of uncertainty and the distributional heterogeneity of return on crude oil.

This research provides a fresh perspective on understanding the risk and return characteristics of crude oil stocks in volatile market environments. It considers distributional heterogeneity and asymmetry of independent variables, offering a new perspective on how these uncertainties affect the ROC collectively. The study also evaluates the indirect effects of pandemics on COP due to macroeconomic changes. The multifactor analysis on ROC is relevant, considering macroeconomic factors like GPR, WUPI, and VIX. The asymmetric quantile approach allows for more flexible and nuanced analysis. The study can help calculate potential return on investment for refinery projects under various market situations.

---

[1] There is evidence to believe that efficient market hypothesis has been the failure in most of the energy market (Liu & Lee, 2018).

The following sections of the study will detail the research methodology and data sets, followed by a discussion of the key empirical findings in the literature. In Section 5, the various statistical tests are covered. The discussion and empirical findings are provided in Section 6.

## 2. Literature review

A growing body of academic work on the relationship between macroeconomic factors and the ROC exists (e.g., McMillan et. al., 2021; Aravind & Nayar, 2019; Hamdi et. al., 2019 etc.). Some studies have used time-varying asymmetric quantile regression methods to examine this relationship (e.g., Dawar et. al., 2021; Xiao & Wang, 2022; Mokni, 2020 etc.), while others have used different statistical approaches like Markov Regime Switching, Vector Auto Regression etc. (e.g., Mahmoudi & Ghaneei, 2022; Golitsis et. al., 2022 etc.).

Several recent studies have examined macroeconomic factors in the context of ROC, e.g., Aravind & Nayar (2019) and Bredin et al. (2021). Macroeconomic conditions and the price of oil have a long-term dynamic relationship (Aravind & Nayar, 2019). The important work of Fama, 1990 demonstrated that INDPRO had a beneficial impact on future cash-flow and thus market returns. These studies have contributed to the literature by providing insights into the underlying dynamics of the relationship between COP and macroeconomic variables. However, most of the studies have been limited to 2-factor or 4-factor analysis (e.g., Cedic et, al., 2021, Shahzad et. al., 2021, Bahloul & Amor, 2021, McMillan et. al., 2021) with few studies using more comprehensive models introducing more factors (e.g., Zhang & Hamori, 2022, Ghosh, 2022, Aravind & Nayar. 2019, Zhang & Hamori, 2022). However, macroeconomic factors are not the only drivers of ROC, there are many other external factors such as GPR, WUPI, GEPU etc. that can directly or indirectly influence the ROC.

Researchers have investigated GPR to examine the effects of global tension, friction, and conflict on the oil-stock markets associations (e.g., Antonakakis et. al., 2017; Wang et. al., 2021; Plakandaras et. al., 2019). Researchers have shown that unexpected and natural events such as pandemics can impact investors' sentiments and affect risk-taking behavior (Shaikh, 2022; Kaplanski & Levy 2010). GPR is a significant indicator that can contribute to a climate of uncertainty and affect economic performance and asset markets (Drakos & Kallandranis, 2015, Schneider & Troeger, 2006). Indeed, the oil market index can be severely affected by the GPR & WUPI but mostly reported short-term instability. Even though stock market responses to global crises are frequently unfavorable, GPR can offer useful data regarding oil volatility and can offer the greatest potential for financial benefits (Liu et al., 2019). Crude oil prices are expected to rise in an atmosphere where there is both market instability and inflation. Oil can become more expensive for buyers using a currency whose value has fallen due to inflation. Increased oil prices can also be a result of choppy markets, which are marked by swings and uncertainty. This is because traders and investors may be more willing to pay more for a commodity in an unstable market.

Economic policy uncertainty (GEPU) can create uncertainty in the market and affect the demand for crude oil (Lei et. al., 2019). This increased market volatility could create opportunities for investors to buy and sell crude oil as prices fluctuate. Furthermore, as investors look to safeguard their money from economic uncertainty, GEPU may result in an increase in demand for crude oil as a safe-haven asset which could result in greater crude oil prices and earnings (Olayeni et. al., 2020; Mensah et. al., 2017). Our study aims to examine the combined impact of a range of factors on ROC, which can provide valuable information for risk management and portfolio optimization.

An increasing corpus of research suggests non-linear framework between oil prices and economies, despite the studies' primary focus on linear models (e.g., Salisu et. al., 2019; Pan et al., 2017; Le and Chang, 2013). According to Beckmann & Czudaj (2013), nonlinearities may result from significant oil price shocks brought on by external variables, discrete regime transitions, or the fundamentally nonlinear nature of the technique used to generate the data (Alqaralleh, 2020). There is no agreement on the most effective methods for performing multifactor analyses and diagnostic tests for WTI excess returns in the literature that is currently accessible, which may be the cause of the existing literature's lack of specificity. We also suggest that a moderation arises when two variables interact in a way that considers the moderating impact, in line with statistical theory. Our current work improves the ongoing discussions in the relationship between ROC, WUPI, GPR and INFLATION by introducing the interaction term.

There are gaps in the literature when it comes to understanding the risk and return characteristics of crude oil stocks in relation to macroeconomic variables, WUPI, GPR, and GEPU, even though there have been extensive empirical studies devoted to the relationship between oil prices and macroeconomic variables. While previous studies such as

McMillan et. al. (2021), Aravind & Nayar (2019) and Hamdi et. al. (2019) focused on the relationship between oil prices and macroeconomic variables, and how they affect the volatility of oil prices, they have not fully explored the implications of these findings on the risk and return characteristics of crude oil stocks. This gap in the literature is significant because understanding the risk and return characteristics of crude oil stocks is important for investors, as it can inform investment decisions and portfolio construction. Additionally, understanding the factors that drive the return of crude oil stocks and how they vary across different quantiles of the return distribution can provide valuable insights into the underlying dynamics of the oil market and potential risks and opportunities.

Our study uses SQR to examine the relationships of the above discussed variables to close this gap in the literature. This unique research offers a fresh viewpoint on how to comprehend the risk and return characteristics of crude oil stocks in various market environments. This study aims to build upon the work of Fung & Hsieh (2004) and Jurek & Stafford (2015) of the methodological difficulties in applying conventional models by using a multivariate approach to analyze the relationship between COP and various macroeconomic variables, global pandemics WPU, GPR, and GEPU. By considering these additional factors our study aims to provide a more comprehensive and accurate understanding of the factors that drive the ROC and how they vary across different quantiles of the return distribution. This can provide valuable insights into the underlying dynamics of the oil market and potential risks and opportunities.

## 3. Model and Econometric Approach

The SQR measures both upper and lower tail reliance in addition to the average or linear dependence between the variables. As a result, the results of the effect of conditional variables on the dependent variable are more exact and precise (Koenker & Ng, 2005). The model can be mathematically formulated (Eq. (1)) considering y be a dependent variable that is assumed linearly dependent on x.

$$Q_y\left(\frac{\tau}{x}\right) = \inf\left\{\frac{b}{F_y\left(\frac{b}{x}\right)} \geq \tau\right\} = \sum_k \beta_k(\tau)x_k = x'\beta(\tau) \qquad \ldots\ldots\text{Equation 1}$$

The conditional quantile function of the response variable (the ROC) is modelled in Eq. (1) as a function of the predictor variables. Here, a conditional probability distribution function is present for y; given x noted by $F_y\left(\frac{b}{x}\right)$ and $\tau \in (0,1)$. The division of the endogenous variable into proportions below and above occurs at this step. Because of this, the structure of reliance is defined as $\tau(x_k)$ do not differ between quantiles; expanding (declining) if they do; and similar (dissimilar) for low and high quantiles indicating an asymmetric (symmetric) dependency (Koenker, 2005). The coefficients $\beta(\tau)$ can be computed as follows by minimizing the weighted absolute differences between y and x:

$$\hat{\beta}(\tau) = \arg\min \sum_{t=1}^{T}\left(\tau - 1_{\{y_t < x_t\beta(\tau)\}}\right) |Y_t - x'_t\beta(\tau)| \qquad \ldots\ldots\text{Equation 2}$$

where $\{y_t < x_t\beta(\tau)\}$ is the usual indicator function.

The multifactor market model can be estimated to determine the oil beta. The regression equation has the following form (Eq. (3)) since the model in this case has a factor intensity structure.:

$$E(r_t) = \alpha + \beta_M R_M + \beta_{t\_1} * F_1 + \beta_{t\_2} * F_2 + \ldots + \beta_{t\_n} * F_n + \varepsilon_t \qquad \ldots\ldots\text{Equation 3}$$

In Eq. (3) $E(r_t) \rightarrow$ Excess return, $\alpha \rightarrow$ intercept, $t \rightarrow 1,2 \ldots, T$, $\beta \rightarrow$ coefficient which is the sensitivity of the asset (WTI) in relation to the specified factor, $R_M \rightarrow$ excess market return on the stock, $F \rightarrow$ systematic factor, $\varepsilon_t \rightarrow$ the asset's idiosyncratic random shock with mean of zero. This is calculated using SQR with an estimate of conditional median (0.5 quantile) and model adequacy was checked using various regression diagnostic tests.

### 3.1. Data source and variables

The study aims to investigate the association between the excess return on spot price of WTI as a proxy for the oil price and macroeconomic indicators. It is important to realize that the period under study, 2017 to 2022, was a particularly turbulent time for the global stock market, marked by the outbreak of a pandemic and escalating geopolitical tensions. Notably all the stock returns exhibit a significant amount of kurtosis and skewness which implies that skewness or kurtosis may be another risk factor.

To estimate betas, five years of monthly data of the selected variables were collected from FRED Economic Data[2]. The selection of variables and data sources are summarized in Table 1. This study uses yield data for the crude oil data as it represents the changes in intrinsic value and provide additional insights. Following consideration of the representativeness, transparency, and consistency of the crude oil data, the WTI crude oil spot price[3] was chosen as the crude oil price variable. The SPY index tracks the S&P 500 index, which is made up of 500 large- and mid-cap US stocks and acts as one of the main benchmarks of the US equity market. The SPY[4] is utilized in this article to both symbolise the financial stability and health of the US economy and to capture shocks to the US stock market. Table (1) presents selected variables for the study.

Table 1. Variable selection & data source

| Factors | Characterization variable | Abbreviation |
| --- | --- | --- |
| U.S. Treasury Securities at 3-Month Constant Maturity | DGS3MO index | DGS3MO |
| U.S. Treasury Securities at 5-year Constant Maturity | DGS5 index | DGS5 |
| Industrial Production: Total Index | INDPRO index | PROD |
| Consumer Price Index for All Urban Consumers | CPIAUCSL | INFLATION |
| Unemployment rate | UNRATE index | UNRATE |
| Narrow money supply | M1SL index | M1SL |
| Change in exchange rate | CCUSMA02EZM618N index | CCU |
| Standard & Poor's Depositary Receipts (SPDR) S&P 500 exchange traded fund (ETF) | SPY index | SPY |
| CBOE Market Volatility Index | VIX index | VIX |
| Geopolitical Risk Index | GPR data | GPR[5] |
| Global price of Energy index | PNRGINDEXM index | GPE |
| World Pandemic Uncertainty Index | WUPI index | WUPI |
| Global economic policy uncertainty | GEPU index | GEPU |
| International crude oil price | WTI crude oil spot price | WTI |

Note: Given S&P's reduction of the nation's credit rating in 2011, the common perception that U.S. treasury securities are devoid of credit risk may be debatable; nonetheless, that subject is outside the purview of this study.

Results show that extreme negative returns tend to be larger in absolute value compared to positive one, which may be attributed to the inclusion of crisis periods such as the ongoing pandemic and GPR during the sampling period. High interest rates can make investments less attractive by reducing the current value of future cash flows. As such economic theory predicts that an increase in interest rates will lead to a result in a decline in stock values. The US treasury bill rate is considered here as risk free interest rate.

*3.2. Econometric approach*

The technique of this study can be divided into two stages: the first stage examines the excess return over time, and the second stage analyses the excess return's cross-section components. Our study's major presumptions are that markets are efficient, events cannot be predicted, and time is affected exogenously.

Eq. (3) can be extended to Eq. (4) to discuss the excess return on WTI.

$$R_{wti} = \alpha_1 + \beta_M R_M + \beta_1 SP_t + \beta_2 SPREAD_t + \beta_3 INDPRO_t + \beta_4 INFLATION_t + \beta_5 UNRATE_t + \beta_6 M1SL_t + \beta_7 CCUS_t + \beta_8 VIX_t + \beta_9 GPR_t + \beta_{10} WUPI_t + \beta_{11} GPE_t + \beta_{12} GEPU_t + \varepsilon_t$$

……Equation 4

Here, $R_{wti}$ = the excess ROC; the risk-free rate which is 3-month US Treasury bill rate here, was deducted from the continuously compounded returns to transform the WTI returns into excess returns, $SP_t$ = the excess market return, $SPREAD_t$ = 5-year minus 3-months treasury yield curve, rest all are as shown in Table (12).

---

[2] FRED data obtained from https://fred.stlouisfed.org/
[3] WTI data obtained from https://www.eia.gov/dnav/pet/pet_pri_spt_s1_d.htm
[4] SPY data obtained from https://finance.yahoo.com/quote/SPY/history/
[5] GPR data obtained from https://www.matteoiacoviello.com/gpr.htm

Since the WUPI effect and the aggravation of GPR have not been evaluated, Model (4) may be underspecified. Therefore, model (2), which augments model (1) by adding the interaction term to the WUPI & GPR nexus, may be more applicable. According to past research (Bodie et al., 2010; Zaremba et al., 2020), Eq. (5) explains the excess ROC in a multivariate framework, and Eq. (6) discusses the relationship between GPR and INFLATION. Table (2) justifies the considered variables.

$$R_{wti} = \alpha_1 + \beta_M R_M + \beta_1 SP_t + \beta_2 SPREAD_t + \beta_3 INDPRO_t + \beta_4 INFLATION_t + \beta_5 UNRATE_t + \beta_6 M1SL_t + \beta_7 CCUS_t + \beta_8 VIX_t + \beta_9 GPR_t + \beta_{10} WUPI_t + \beta_{11} GPE_t + \beta_{12} GEPU_t + \beta_{13} (GPR_t * WUPI_t) + \varepsilon_t$$

……. Equation 5

$$R_{wti} = \alpha_1 + \beta_M R_M + \beta_1 SP_t + \beta_2 SPREAD_t + \beta_3 INDPRO_t + \beta_4 INFLATION_t + \beta_5 UNRATE_t + \beta_6 M1SL_t + \beta_7 CCUS_t + \beta_8 VIX_t + \beta_9 GPR_t + \beta_{10} WUPI_t + \beta_{11} GPE_t + \beta_{12} GEPU_t + \beta_{13} (GPR_t * INFLATION_t) + \varepsilon_t$$

------Equation 6

The interaction term used in Model (2), Eq. (5), to analyze how the impact of GPR on ROC may change during a pandemic, or vice versa. This information can be used to identify potential risk factors for investing in WTI during times of geopolitical upheaval and pandemics, and it can also be used to inform investment decisions. Furthermore, the SQR approach allows for the examination of the effect of these factors on the ROC at various levels of the outcome variable, which can aid in the identification of investment possibilities that may be less affected by these factors. Moreover, in the case of Model (3), Eq. (6), the GPR and INFLATION are two factors that can have an impact on the oil market, and the interaction term allows us to observe how these two factors interact. By examining the interaction term, we expect to have a clearer understanding of how GPR might affect oil prices during periods of high or low inflation.

Table 2. Justification of selected factors

| | |
|---|---|
| SPY | Since SPY is a commonly used benchmark for the U.S. stock market and can be used to determine the state of the economy as a whole, it is seen as a crucial indicator. A booming economy and increased demand for commodities like oil, which can increase the price of WTI, are frequently connected with the S&P 500 index's robust performance. On the other hand, a poor stock market performance may signal a weak economy and decreased demand for commodities, which could pull the price of WTI lower. |
| PROD | Existing work (Alao & Payaslioglu, 2021) suggests that there is a dynamic link between the price of oil and POOD. According to the study, an increase in PROD may lead to higher demand for crude oil, resulting in an increase in oil prices and potentially boosting the ROC. This relationship can provide valuable insights for investors looking to make informed decisions about investments in the oil industry. So, our study has incorporated PROD as a crucial factor for further investigation. |
| M1SL | Controlling money supply has been traditionally considered as the primary responsibility of central monetary authorities in any given economy because it has an impact on economic activity (Osamwonyi, et al., 2012). According to the study, as people, as rational economic agents, tend to diversify their wealth holdings from financial assets (such as stocks and shares) to real assets, if money growth does not match with growth in output of goods and services, it can cause inflationary spiral in the economy and lead to lower stock prices. This highlights the importance of central monetary authorities in maintaining monetary stability and preventing inflation, which can have negative impacts on the stock market. Narrow money supply is taken as crucial factor for ROC. |
| INFLATION | We utilized US inflation rates as a proxy for global inflation rates. According to studies, there is a significant causal connection between oil prices and the CPI. The CPI frequently shows a negative sign of feedback to oil prices, suggesting that higher inflation leads to higher interest rates, which in turn cause the economy to contract and lower oil prices and demand (Tiwari et. al., 2019). |
| WUPI | The ROC has been significantly impacted by the pandemic. The disruption caused by the pandemic on the world economy has led to a decrease in demand for crude oil, which in turn has lowered the price of oil. As a result, investors in crude oil stocks have seen a decline in their ROC. According to the report, the epidemic caused an abrupt reduction in oil prices of 30%, the biggest drop since the Gulf War in 1991. This demonstrates the crude oil market's sensitivity to world events and the possible risks for investors in this area (Prabheesh et al. 2020; Iqbal et al. 2020). |
| GPE | Since it reflects the average price of energy commodities, such as crude oil, globally, the GPE is regarded as a reliable predictor of the ROC. Changes in the GPE on a worldwide scale may influence the supply and demand for crude oil, which may influence the price of WTI. Lower GPE may imply weaker demand, which would put downward pressure on WTI, while higher GPE may indicate robust demand for crude oil, which would push up the price of WTI. |
| VIX | VIX, also known as the "fear gauge" of the market, reflects investor sentiment and predictions about future market volatility. It is commonly observed that when the stock market declines, the change in VIX increases at a higher rate than when the market rises (Chen & Zou, 2015). This is a widely accepted phenomenon in finance, it is also known as the VIX term structure, and it can be used as a measure of market sentiment and risk. |

| GPR | Academics have focused increasingly on the impact of GPR on the dynamics of the stock market in recent years (such as Bouri et al., 2019; Caladara & Iacoviello, 2018). Other researchers have also discovered the influence of macroeconomic indicators on the stock market, such as the GPR, the World Uncertainty Political Index (WUPI), and Industrial Production (INDPRO) (Gong et. al., 2022 and Chien et. al., 2021). This emphasises how crucial it is to consider GPR and other macroeconomic indicators to comprehend stock market trends and make wise investment decisions. The volatility of GEPU and GPR can have a significant impact on the oil market index (Wang et. al., 2022; Antonakakis et. al., 2017). |
|---|---|
| SPREAD | WTI excess return and SPREAD (the Treasury yield curve) changes may influence one another (Idilbi-Bayaa & Qadan 2022). Furthermore, we believe that rising crude oil demand may raise oil prices, boosting excess profits on investments in WTI. This can affect investors' expectations of inflation, which might cause the Treasury yield curve to steepen as a result. On the other hand, if crude oil demand declines, it would result in lower excess returns on a WTI investment and flatten the yield curve. The intricate interrelationship between the crude oil market, inflation, and Treasury yield curve is highlighted here, demonstrating how changes in one aspect can have repercussions on other variables. |
| GEPU | The previous studies that looked at the negative impacts of WUPI and GEPU on the oil market index are cited in this paper. This subject has been studied by Wang et al., 2022, and Antonakakis et al., 2017. Furthermore, Diebold & Yilmaz (2014) discovered in their initial research that crude oil prices react, regardless of the time scale, to information on economic policy uncertainty. This is also reinforced by Lei et al. 2019, who stressed that the market and demand for crude oil might be impacted by the unpredictability of economic policy. This emphasises how crucial it is to consider international political and economic considerations to comprehend the dynamics of the crude oil market and make wise investment choices. |
| CCU | The US dollar (USD) exchange rate is an important variable to consider when analyzing the crude oil market, as the price of oil is often quoted in USD. The study uses the average of daily rates of the USD exchange rate for the Euro Area (19 countries) as a variable in the model. This allows the study to differentiate between the impact of changes in the COP and CCU on the ROC. |
| UNRATE | COP shocks have a significant impact on macroeconomic indicators such as inflation, UNRATE, CCU, and PROD. It also cites evidence that these indicators have a long-term co-integrating relationship with the price of WTI crude oil (Mensi et. al., 2020). This highlights the importance of considering the broader economic context when analyzing the crude oil market and making investment decisions. |

## 4. Method

For each of the variables, a series of modifications must be made. This is based on the Arbitrage Pricing Theory (APT), which contends that unanticipated changes in macroeconomic factors, rather than their levels, can be used to explain stock returns. The study makes the naive assumption that investors' expectations for the future value of the variables will remain unchanged. Thus, the unforeseen change is the overall variation in the variable from one period to the next. Eq. (7) displays the calculation of the monthly logarithmic excess returns for WTI, where the 3- month U.S. Treasury rate is used as the risk-free rate.

$$ER_{wti(t)} = ln\left(\frac{p_{wti(t)}}{p_{wti(t-1)}}\right) - r_f \qquad \ldots\ldots \text{Equation 7}$$

In Eq. (7), $ER_{wti(t)}$ is the excess return of WTI at time $t$, $P_{wti(t)}$ is the price of WTI at time $t$, $P_{wti(t-1)}$ is the price of WTI at time $t-1$, and $r_f$ is the 3-month U.S. Treasury rate. The monthly yield on a three-month U.S. T Bill is subtracted from the continuously compounded daily returns on the WTI Index to determine monthly excess returns. The macro-economic factors, which serve as the predictors, are expressed as the log changes of the data. VIX and SP are expressed in levels. We use the Eq. (8) to calculate the daily log changes:

$$VIX_t = ln\left(\frac{VIX_t}{VIX_{(t-1)}}\right) \text{ and } SP_t = ln\left(\frac{SP_t}{SP_{(t-1)}}\right) - r_f \qquad \ldots\ldots \text{Equation 8}$$

where $VIX_t$ and $SPY_t$ are the daily log change at time $t$, $VIX_t$ and $SP_t$ are the values at time $t$, and $VIX_{(t-1)}$ and $SP_{(t-1)}$ are the values at time $t-1$. Rest all the factors are standardized using (lnDiff($X_t$) - lnDiff($X_{t-1}$)), where X is the respective factor and $X_{t-1}$ is the value at previous time.

Table (7) presents the descriptive statistics for all the variables. The table comprises the variance inflation factor (VIF), the Augmented Dickey-Fuller (ADF) (Dickey & Fuller, 1979) unit root test empirical statistics, and the test statistic for the Jarque-Bera (JB) test for the normality assumption. VIF values for the independent variables are all <3.0 indicating multi-collinearity does not present (Hair et al., 2017). The ADF-tests demonstrate that all series are stationary. Many of the variables show skewness and high amount of kurtosis. This raises the possibility that skewness or kurtosis could represent another risk factor. Market returns and financial time series distributions frequently exhibit skewness and a heavy tail (Xiao et. al., 2015). Negative skewness (Fig. (7)) suggests that more negative data is concentrated on the mean value despite positive mean indicating favourable results on average yield for investors. Additionally, the dependent variable is typically considered to be independently distributed and homoscedastic. Large excess kurtosis coefficients which is leptokurtosis, are a sign that outliers are present and indicates that there have been numerous price changes in the past (either positive or negative) away from the average returns for the investment. Almost 50% of the data set includes the 2020-21 pandemic crisis and a post crisis period, which becomes apparent by observing the large standard deviation levels associated to some of the variables. For all series, the JB test statistics reject the null hypothesis ($H_0$) of a normal distribution at the 5% significance level.

Considering the minimum values, the lowest in this range is the UNRATE with a minimum value of -131.38. GEPU is much dispersed than other variables with a standard deviation of 43.40; closely following this is the GPR with 30.93, UNRATE with 21.29 and MONEY with 20.32. Negative values for skewness are common (SP, CURRENCY, MONEY, UNRATE, INFLATION, GPR & SPREAD) but are positive for the INDPRO, PANDEMIC, GPE, VIX & GEPU. Most of these factors show excess kurtosis. The last thing to be concerned about in this type of multifactor modelling study is the incidence of multicollinearity. The variance inflationary factors (VIF) are displayed; in no case does the VIF for any of the factors even come close to the critical value (VIF > 5) (Hair et. al., 2017). This suggests that multicollinearity, while present, is not too much of a problem. But to develop a new coordinate system and align it with the largest variation in the data, Principal Component Analysis (PCA) was carried out. The results displayed in next section.

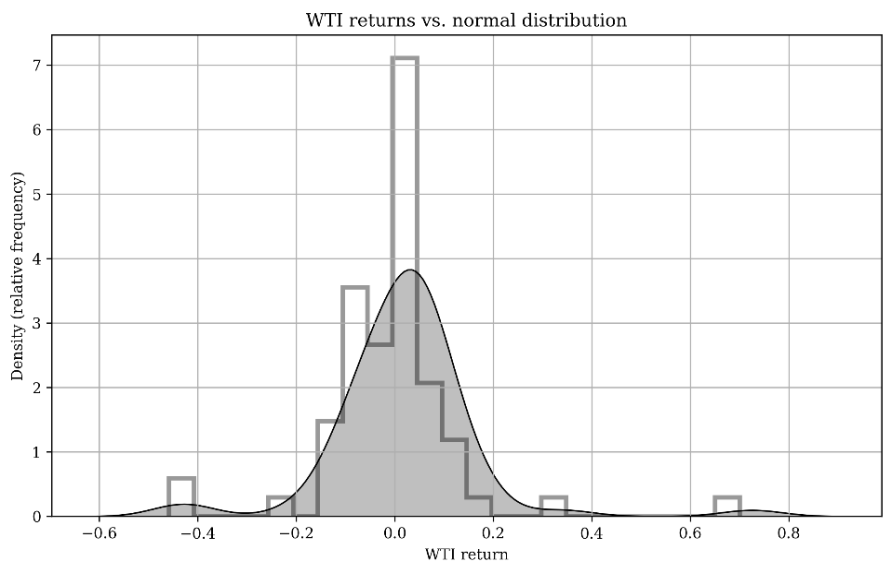

Figure 1. Skewed target distribution

Value at Risk (VaR) was estimated (Table 3) on simple returns which represents the worst-case loss associated with probabilities and Cvar was estimated by averaging the severe losses in the tail of distribution of WTI returns.

Table 3. WTI Value at Risk

| Confidence level | VaR | Conditional VaR |
|---|---|---|
| 90% | -0.10 | -0.22 |
| 95% | -0.13 | -0.30 |
| 99% | -0.43 | -0.43 |

Quantile normalisation process was performed to alter the raw data to preserve the genuine variation that we were interested in investigating while removing any undesired variation caused by technological artefacts. Fig. (2) displays the normalized box plot of the dependent variables.

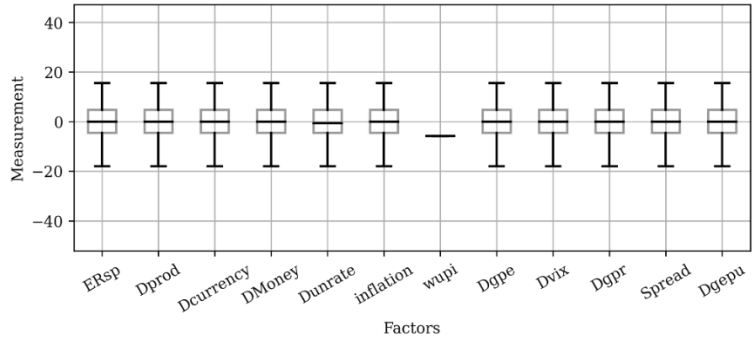

Figure 2. Boxplot of data after Quantile Normalization

The proportion of eigenvalues attributed to each component is shown in Fig. (3). This indicates the importance of each component for the analysis.

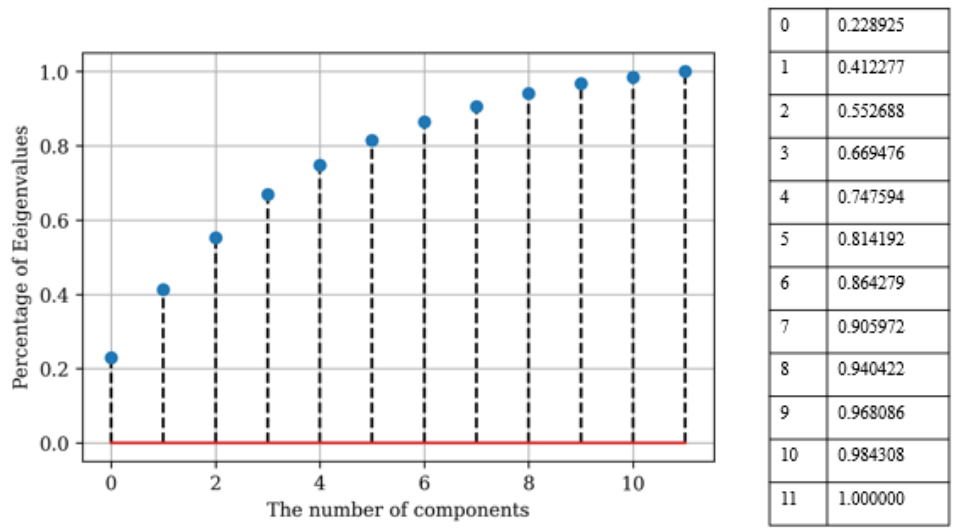

Figure 3. Percentage of Eigenvalues Attributable to Each Component

## 5. Multifactor Quantile estimates

Tables (8-10) report the estimates of the SQR for the ROC. The distributions divided into nine different quantiles (i.e., $\tau = 0.10 – 0.90$) to get a mixed variety of low, medium, and high return conditions. Values of that are too close to its limits of 0 and 1 do not usually have a good match, hence these values were avoided in this analysis. Numerical results are displayed with consideration of the WUPI and GPR. The OLS regression line and the regression line for q=50 percent are identical. Table (8) reports the regression estimation ($Q_n0.5$) based on Eq. (24). The diagnostics tests were performed on conditional median quantile which has been treated here as the estimation results for the baseline regression. The asymmetry in the model can be noticed by contrasting the coefficients of various quantiles.

A few parameter estimations, notably those for the "dCURRENCY", "dUNRATE", "dWUPI", "dGPE", "dGPR", and "dSPREAD" variables, are not statistically distinct from zero. The F-test, which adds the predictive power of all independent variables and demonstrates that it is implausible that all the coefficients are equal to zero, was used to test the $H_0$ that the parameters for these six variables are all zero. These variables do not seem to perform much better or worse than the WTI stock, either. The $H_0$ that the estimate is different from 0 cannot be disregarded, as shown by the F-test statistic value of 0.911 and the p-value of 0.494. All the variables taken into consideration have a sufficient impact on ROC, as evidenced by the rejection of p-value. However, F-value 0.911 < critical value 3.44 (Pesaran et al., 2001 lower bound critical value) at 5% significance level implies that $H_0$ cannot be rejected and there does not existing any long run relation with COP and these variables.

The heteroscedasticity is assessed using the Breusch-Pagan test. Here, homoskedasticity is assumed by the $H_0$. Therefore, we reject the $H_0$ and conclude the existence of heteroskedasticity if p val 0.05. The heteroscedasticity provides the justification to examine the function in different quantiles. The results show Lagrange multiplier statistic (41.56), p-value (0.00), f-value (7.51), and f p-value (0.00). p-values for both being < 0.05 indicating fundamental problem of heteroscedastic errors. Fig. (3) displays the residual plot, though any clear pattern is not visible, however, Jarque-Bera (JB) normality assumption test was performed to ensure the correctness of our assumption. According to Fig. (4), the pandemic caused an early decline in prices throughout 2020-21, followed by a steep rise as producers reduced supply and demand soared.

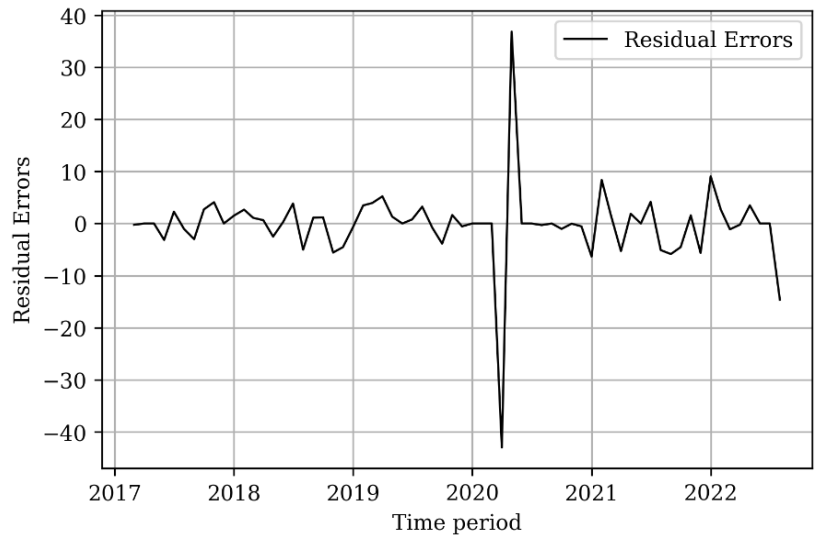

Figure 4. Patterns in the residuals over time

The assumption is satisfied because the Durbin Watson's test result of 1.98 indicates that there is no autocorrelation. Following that, a normality test was run on the residuals, with the premise that the model's residuals are normally distributed. Table (4) reports the normality test.

Table 4. Jarque-Bera normality test

| Jarque-Bera | 3452.12 |
|---|---|
| Chi² ($\chi^2$) two-tail prob. | 0.00 |
| Skew | -5.19 |
| Kurtosis | 36.87 |

The statistic and $\chi^2$ two-tailed p-value for the test that both the kurtosis and the skewness are consistent with a normal distribution, which is to say that the residuals are overall normally distributed.

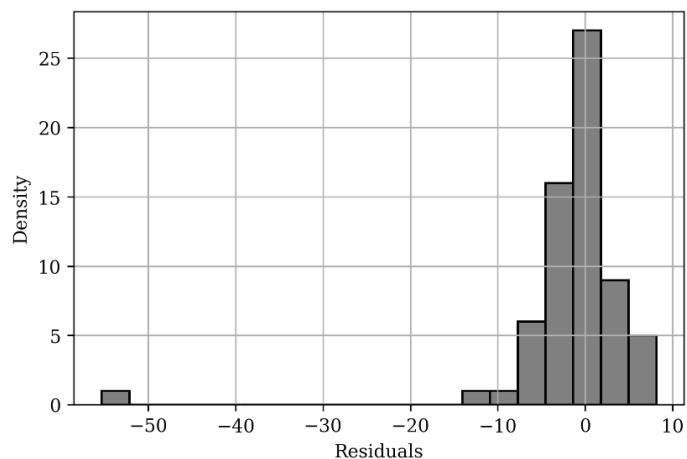

Figure 5. Histogram of residuals

According to the histogram plot in Fig. (5), the distribution of the residuals is bell-shaped; nonetheless, there are several substantial negative outliers that could lead to a significant negative skewness. It appears that a limited number of big negative residuals, which exhibit monthly WTI price declines of greater than 15% and most recently 60%, are what are to blame. Fig. (6) displays the regression residuals and fitted series. Numerous significant (negative) outliers may be seen in the graph, but the largest one is in 2020.

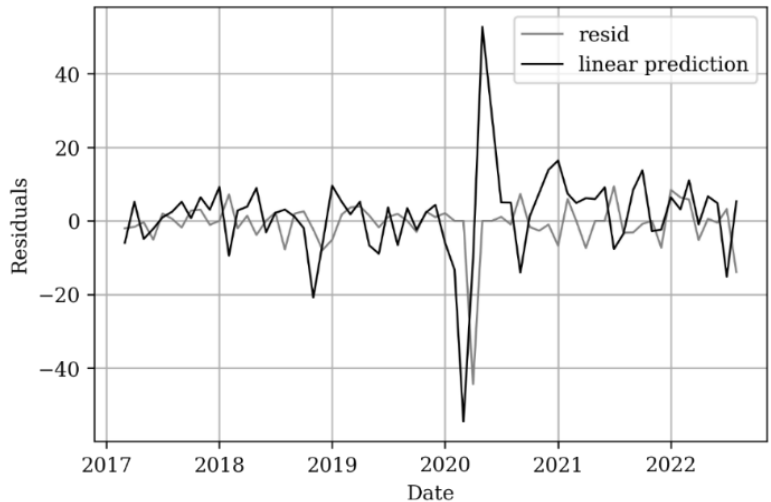

Figure 6. Regression Residuals and Fitted Series

A table of values for the residuals was studied to determine the precise dates when the largest outliers were realised. 1st dummy variable was added to explain the COVID outbreak and lockdown effect, 2nd dummy variable was added to explain Ukraine war.

Table 5. Dummy variables construction.

| Date | Smallest residuals |
|---|---|
| Dummy variables with exogeneous variables | |
| 2020-04-01 | -55.36 |
| 2022-08-01 | -12.96 |
| Dummy variables with interaction term (WUPI * GEUPI) | |
| 2020-03-01 | -74.29 |
| 2022-08-01 | 21.45 |
| Dummy variables with interaction term (GPR * INFLATION) | |
| 2019-01-01 | -44.68 |
| 2022-05-01 | -28.42 |

It is evident from Table (5) that the two most extreme residuals were in April 2022 (– 55.36) and August 2022 (– 12.96). Due to the perfect fit of the dummy variables to the two extremely outlying observations, the rerun of the regression along with the dummy variables significantly increases the pseudo $R^2$ value from 0.58 to 0.71. The parameters of the highly significant dummy variables in the model correspond to the levels that the pertinent residuals would have attained if the dummy variables had not been employed.

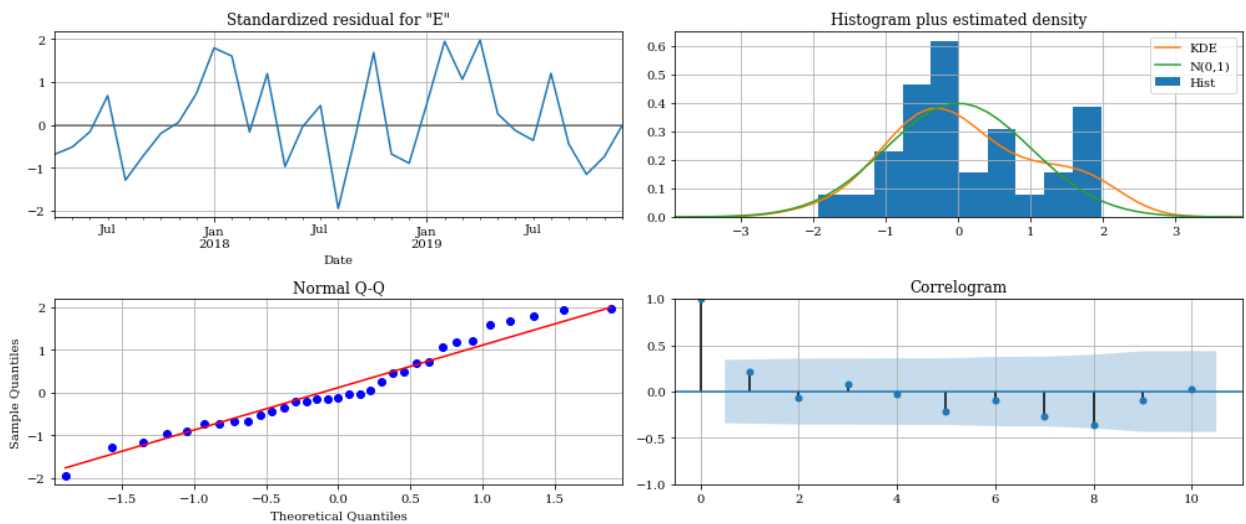

Figure 7. Residuals diagnostics

Fig. (7) displays the residuals plot where it can be observed that the errors follow a normal distribution. This has effectively established a baseline model to estimate the effect of the event on our target variable. Furthermore, both missing variables and inappropriate functional form were discovered using the RESET. An F-value of 0.008 and a corresponding p-value of 0.9251 from the data show that we cannot rule out the $H_0$ that the model contains no omitted variables. In other words, there is not anything to suggest that the model's functional form of choice is flawed. To ascertain whether there is a structural break in the data at any given moment, the CUSUM test (Ploberger & Krämer, 1992) for parameter stability based on OLS residuals was carried out. It is common for the date of the structural break to be unknown in advance. The CUSUM non-parametric method tests for the presence of a change at each possible point in the data rather than specifying the exact date of the change. Table (6) present the cumulative total and cumulative sum of squares of recursive residuals to test the structural stability of the models. The absence of any structural breakdowns is the null hypothesis.

Table 6. Parameter stability test

| test statistic | 1.96 |
| --- | --- |
| p-value | 0.018 |
| Critical values | [(1, 1.63), (5, 1.36), (10, 1.22)])] |

Based on Table 6's test statistic and associated p-value, the H0 that the coefficients are stable through time can be rejected because our model does indeed contain a structural break for each break date in the data.

*5.2. Casual impact analysis*

In the time following the intervention, the response variable's average value was 1.36. Without the intervention, we would have anticipated a 3.21 average response. The response variable had an overall value of 43.6 when the post-intervention period's individual data points were added together. But if the intervention had not happened, we would have anticipated a total of 116.77 in absolute terms, with a confidence interval of [80.29, 154.44]. With an upper and lower bound of [-94.96, -31.46], the response variable showed a relative decline of -62.7%. This demonstrates that the detrimental impact seen during the intervention period is statistically significant. Fig. (8) displays the casual impact analysis plot. The Bayesian one-sided tail-area probability of getting this result by chance is exceedingly low (p = 0.0). This indicates that the causal effect is statistically significant.

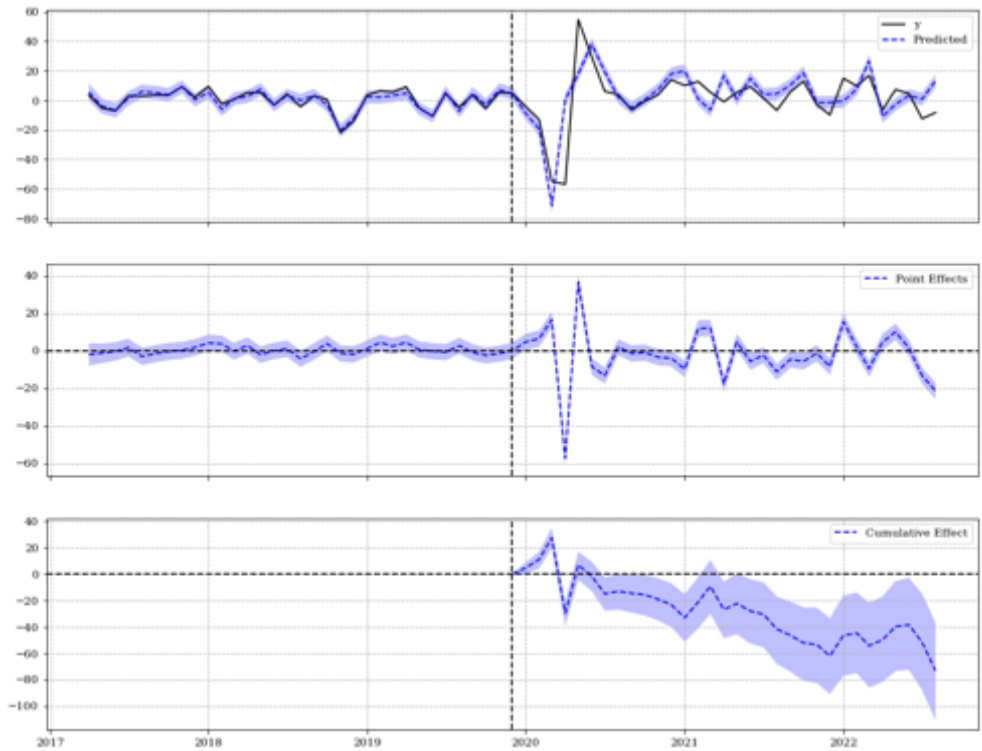

Figure 8. Casual Impact plot

## 6. Empirical results & discussions

The quantile analysis found the following intriguing trends. First, the predicted coefficient is significant across all quantiles; it is negative in lower quantiles across the board for the entire model but positive in upper quantiles. With an upper tail dependence and a lower tail independence, this shows that the dependence structure is asymmetric.

The coefficient on the intercept term is positive and statistically significant for the variables SP, PROD, MONEY, UNRATE, INFLATION, WUPI, GPE, VIX, and GEPU at 5% level implies that substantial impact on ROC when the market is bullish; however, negative coefficient for SPREAD, CURRENCY and GPR at the very lowest quantiles implies a prolonged drop in investment during bearish market. However, the negative effect is not statistically significant at 5% level. Table (7) presents a complete discussion on each factor.

Table 7. Empirical estimation

| SPY | Regarding the control variables, the effect of SPY stocks on WTI stocks is favourable as anticipated. It is positive and significant for all quantiles except insignificant for WUPI and GPR interaction. Such findings are confirmed by Hung & Vo, 2021 and Pal & Mitra, 2019. However, the combined effect of WUPI & GPR seems to have severe implications on COP volatility and this combined effect is higher than on the GEUPI (Sharif et. al., 2020). |
|---|---|
| PROD | A positive and significant coefficient for the INDPRO variable in the upper tail of a ROC suggests that as industrial production increases, ROC are more likely to be in the upper tail (i.e., above the median return). This implies of a strong industrial sector is associated with higher ROC. |
| CCU | A change in CCU that is not statistically significant across quantiles but mostly shows positive signs indicating, while a change in CCU may have some effect on ROC, it is not a strong determinant. The positive signs suggest that as the CCU changes, the ROC are more likely to rise, but this relationship is not strong enough to be considered definitive. |
| M1SL | A significant and positive sign for MISL across all the quantiles indicates that as the M1SL expands, so will the ROC. This relationship holds true across all quantiles, implying that as the M1SL expands, ROC are likely to rise regardless of whether they are in the lower, middle, or upper part of the distribution. This suggests that the M1SL is a significant determinant of ROC. Asymmetric relationships of Oil Prices, Money Supply have been studied by Bin & Rehman, 2022 and they have found long-run and short-run relationships. However, no studies have relay emphasized the importance of M1SL for ROC perspective. |
| UNRATE | The fact that the UNRATE varies in significance across quantiles suggests that the relationship between the UNRATE and the ROC is not consistent across all levels of the return distribution. The varying significance |

| | across quantiles suggests that the relationship between UNRATE and ROC is not straightforward and may be influenced or conditional on other factors. |
|---|---|
| INFLATION | INFLATION being significant with a positive sign across all quantiles in a quantile regression for ROC implies that as inflation rises, ROC are likely to rise as well. This relationship holds true across all quantiles, implying that as inflation rises, ROC will rise as well, regardless of whether they are in the lower, middle, or upper part of the distribution. This suggests that inflation is a significant determinant of ROC and can be interpreted as a sign that high inflation will lead to high oil prices, which will result in ROC. This was expected because, COP tend to rise in tandem with inflation, crude oil stocks may serve as an inflation hedge. |
| WUPI | The fact that WUPI is significant with a positive sign in the quantiles (2-4) suggests that as pandemic uncertainty increases, ROC are more likely to be higher in these quantiles, but this relationship may not hold true for the lower or upper quantiles. WUPI uncertainty being significant with a negative sign in the lower quantiles (2-4) implies that as WUPI uncertainty increases, ROC are more likely to be lower in the lower quantiles (2-4), but this relationship may not hold true for the upper quantiles. The WUPI uncertainty captures the uncertainty about the global economy's future, which can lead to market volatility and lower ROC. When an interaction term between GPR and INFALTION (Model (3)) is included, the WUPI is mostly insignificant, implying that the inclusion of the interaction term has changed the relationship between WUPI and ROC. It could imply that the relationship between WUPI and ROC is dependent on the level of GPR and INFLATION, and that the relationship is insignificant when these factors are considered. |
| GPE | GPE is positive and significant in upper quantile (Model (1)) which signifies when the GPE is high, WTI stock investors can expect higher returns for the top half of the distribution. This also indicate that WTI is sensitive to the global energy market, and that when energy prices are high, WTI stock returns are expected to rise. When the WUPI and GPR interaction term (Model (2)) is included in the model, the GPE has a positive and significant impact on the across all quantiles. This implies that the relationship between GPE and ROC is influenced by WUPI and GPR. It is likely that during times of high pandemic and geopolitical risk, the GPE becomes a better predictor of ROC regardless of where the return falls in the distribution. With the added interaction term between GPR and INFLATION, the GPE is insignificant across all the quantiles which implies that, the relationship between GPE and ROC may be overshadowed by the effects of GPR and INFLATION. |
| GEPU | GEPU has a negative but significant impact on ROC in the upper quantiles, implying that as economic policy uncertainty increases, WTI stock returns become more volatile, showing upper tail reliance. This implies that the negative impact of economic policy uncertainty is more pronounced for higher WTI stock returns. The empirical findings, however, show that the effects of COP shocks and GEPU are asymmetric and closely tied to market circumstances when the interaction terms are considered. Similar findings have reported by You et. al., 2017; Xiao & Wang, 2022). |
| VIX | The VIX has a significant impact on ROC across the entire return distribution. This means that crude oil returns are strongly correlated with stock market volatility, and that when the stock market is volatile, crude oil returns are likely to be lower. This could be because changes in stock market volatility are frequently driven by changes in economic conditions or investor sentiment, both of which can impact demand for crude oil and thus its price. Furthermore, the presence of a significant relationship across all quantiles in all 3-models implies that the relationship between VIX and ROC is consistent and not restricted to specific segments of the return distribution. Previous studies indicate that, VIX showed the least amount of information disturbance before and during the epidemic on all scales (Lahmiri & Bekiros, 2020). |
| GPR | GPR showed no significant relation across any quantile which implies that GPR does not affect ROC in a meaningful way. This could be because the crude oil market is immune to geopolitical events, or because the events considered in this study have little impact on the crude oil market. It is also worth noting that the lack of a significant relationship does not necessarily imply that there is no relationship at all; it could mean that the sample size is too small to detect a relationship, or that the events used as proxies for geopolitical risks do not capture the full picture of geopolitical risks. Earlier studies have reported that Geopolitical uncertainties have a short-term impact on oil prices, lasting less than a year (Jiang et. al., 2022). When the WUPI and GPR interaction term is included in the model, geopolitical risk has a significant impact on ROC at the lower end of the return distribution. This implies that when the WUPI is high, GPR has a greater impact on ROC, particularly at lower returns. This means that when there is a high level of global uncertainty, the crude oil market is more vulnerable to geopolitical events, resulting in lower returns. The inclusion of an interaction term in the model may have helped to capture a more nuanced relationship between GPR and ROC that was not evident when only one variable was used. |
| SPREAD | The spread of Treasury bond interest rates, has no significant impact on ROC across the entire return distribution. This could be because the crude oil market is immune to changes in interest rates, or because the spread of Treasury bond interest rates used in the study does not capture the full range of factors influencing interest rates and thus has little impact on ROC. Furthermore, the lack of a significant relationship across all quantiles suggests that the relationship between SPREAD and ROC is inconsistent or limited to specific segments of the return distribution. |
| WUPI*GPR | In the lower quantiles (0.1-0.5), there is a significant effect indicates tail dependence and a high risk of large losses. This also indicate non-linear relationships and given the investor sentiment at the time, this suggests that there is significant investor pessimism on the likelihood of dropping market prices. |
| GPR* INFLATION | Significant interaction terms are present at the extreme lower tail (Q1.0 & 2.0) of the distribution, and the intercept values are mostly negative, indicating strong relationship with ROC in the lower tail of the distribution. This could imply that the independent variables and the response variable have a non-linear relationship. This set of findings suggests that the lower tail of the distribution is at high risk of experiencing extreme negative outcomes (e.g., large losses or large claim sizes). However, from risk management perspective, this implies a short-term risk measurement and investor's sentiment remain bullish given the time-period (Naifar et. al., 2020). |

Our research has mentioned the essential topic of how ROC and GPR interact against the perilous backdrop of the pandemic. This study found a bearish impact of the pandemics and GPR on the performance of COP which opens opportunity to invest. Our study supports that WTI crude oil can be a cheap hedging tool and when investing a small or larger part in WTI crude oil future market, high hedging effectiveness could be achieved (Dai & Zhu, 2022). Our study also confirms the earlier work of dynamic hedging results which suggest that crude oil futures can provide a profitable hedging opportunity in combination with green energy index (Ahmad, 2017). The most important finding of this exercise is that the WUPI & GPR impacts the crude oil stocks differently. The changes in investor opinions on crude oil stock investments are offered by differences in performance among quantiles. Thus, the WUPI & GPR has not influenced the expectations of investors and this finding confirm the work of Lahmiri, & Bekiros, 2020.

Several macroeconomic factors, including global supply and demand, geopolitical developments, and currency fluctuations, have an impact on the COP. For instance, rising demand for oil from quickly industrialising nations like China and India may push up prices, while falling demand brought on by a recession in a big oil-consuming nation like the United States may push down prices. COP can also be impacted by things like monetary policy changes, natural disasters, and political unrest in nations that produce oil. Considering this, investor expectations, which quickly alter in response to any information made accessible to the public, including economic and political developments and leads to a considerable impact on stock prices. As a result, when examining influences on stock prices, the factors that influence macroeconomy, may be less appropriate to use than those that track changes in expectations about future values of macroeconomic factors. This could be the future direction of this study. Given that investors cannot anticipate or take steps to protect themselves from such risks, indicators that represent unanticipated changes in future macroeconomic variable values are particularly crucial. There are many potential variables that could be taken into consideration because economic theory does not stipulate which parts or how many should be employed in the study.

Our empirical findings have implications for portfolio design and risk management for investors. It also has significant implications for risk management decisions involving hedging and downside risk, given that the financial utility of oil varies depending on market conditions. Finally, our findings have implications for the forecasting of COP across quantiles based on macroeconomic and financial variables. Furthermore, changes in the several parameters taken into account for this study account for almost 2/3 of the monthly fluctuation in the excess returns.

**Conclusion**

This analysis uses multiple factors to model the ROC under different market conditions, considering the impact of various economic, political, and health-related factors on the price of crude oil. Instead of focusing only on the mean or overall trend, this study used SQR to determine how these factors affect the various percentiles (or quantiles) of ROC. In doing so, the study used SQR to evaluate the stability of the relationship between the dependent and independent variables over time, as well as to identify any changes in the relationship that may have occurred as a result of changes in the economic or geopolitical landscape. Furthermore, the study used interaction terms with WUPI*GPR and GPR*INFLATION to conduct additional empirical research. The multivariate 12-factor approach helped to estimate the conditional quantiles of the return distribution, which provides valuable information for risk management and portfolio optimization. The model tested various statistical assumptions e.g., VIF, PCA to satisfy the revenant predictors, Breusch-Pagan test for heteroscedasticity, Jarque-Bera normality test and finally RESET to test structural break in the model. This conclusion is based on the sample of data used in the study and it is possible that results might differ with different data set. The COP is an indicator for world economic development can be viewed as a crucial index for investors and policymakers; thereby, the findings from the study have wider ramifications for both policymakers and investors at large. The asymmetric and heterogeneous association between the given variables indicate that financial specialists and policymakers should embrace distinctive investment strategies under changing economic conditions. Despite the complexity of estimating multiple factors, multifactor SQR has been shown to be beneficial in determining the ROC because it allows for a more comprehensive and accurate analysis of the ROC investments by considering the impact of various economic, political, and health-related factors on COP. Furthermore, the results of this analysis can be used to create a predictive model for forecasting COP under various market scenarios. This can aid in the identification of profitable investment opportunities and the formulation of strategic investment decisions. To this end, building a trustworthy empirical model requires iteration and is not a precise science. Other authors could get a different final specification using the same facts and initial theory.

Table 8. Descriptive statistics of transformed variables

| Variable | Mean | Median | Max | Min | Std Dev | Skew | Kurtosis | JB | ADF | VIF |
|---|---|---|---|---|---|---|---|---|---|---|
| WTI | 0.75 | 3.38 | 54.55 | -56.82 | 14.59 | -0.98 | 8.10 | 161.13 ** | -6.76** | 2.60 |
| SPY | 0.90 | 1.58 | 6.29 | -21.04 | 3.95 | -2.88 | 14.00 | 543.34 ** | -7.07** | 2.29 |
| PROD | 0.01 | -0.03 | 15.75 | -10.31 | 2.76 | 2.16 | 18.51 | 848.54** | -6.36** | 1.95 |
| CURRENCY | 0.00 | 0.00 | 0.03 | -3.21 | 2.01 | -0.11 | 0.49 | 6.49* | -5.46** | 2.14 |
| MONEY | -0.01 | 0.02 | 110.79 | -120.24 | 20.32 | -0.71 | 32.40 | 2466.58** | -6.35** | 2.10 |
| UNRATE | -0.01 | -0.01 | 97.74 | -131.38 | 21.29 | -2.16 | 28.83 | 1955.82** | -6.52** | 2.61 |
| INFLATION | -0.00 | 0.01 | 0.67 | -0.11 | 0.34 | -0.70 | 2.69 | 22.58** | -5.73** | 2.61 |
| PANDEMIC | 0.00 | 0.00 | 4.45 | -2.89 | 0.85 | 1.75 | 14.19 | 501.14** | -7.93** | 1.59 |
| GPE | 0.16 | 1.09 | 46.32 | -0.33 | 11.41 | 0.26 | 4.31 | 42.66** | -7.61** | 3.59 |
| VIX | 0.03 | -0.01 | 2.25 | -0.46 | 0.32 | 4.90 | 32.69 | 2757.41** | -6.77** | 1.29 |
| GPR | 0.06 | 2.41 | 91.15 | -130.65 | 36.32 | -0.41 | 1.48 | 6.34* | -4.74** | 1.83 |
| SPREAD | -0.02 | -0.00 | 0.51 | -0.98 | 0.19 | -1.56 | 9.22 | 222.02** | -5.21** | 1.18 |
| GEPU | 0.83 | -3.34 | 140.48 | -101.33 | 43.40 | 0.63 | 1.48 | 8.78* | -6.10** | 2.14 |

Note: * and ** indicate the rejection of the null hypothesis at the 1% and 5% levels

Table 9. Parameter estimate of Macro-Economic factors on ROC

| Variable | Qn (0.1) | Qn (0.2) | Qn (0.3) | Qn (0.4) | Qn (0.5) | Qn (0.6) | Qn (0.7) | Qn (0.8) | Qn (0.9) |
|---|---|---|---|---|---|---|---|---|---|
| erSP | 0.66 **(0.14) | 0.64 **(0.10) | 0.56 **(0.08) | 0.52 **(0.08) | 0.44 **(0.07) | 0.49 **(0.07) | 0.51 **(0.06) | 0.61 **(0.09) | 0.48 **(0.11) |
| dPROD | 0.39(0.56) | 0.42(0.38) | 0.33(0.31) | 0.69 *(0.3) | 0.82 **(0.27) | 0.9 **(0.29) | 1.0 **(0.22) | 0.89 *(0.43) | 1.35 **(0.23) |
| dCURRENCY | -0.22(75.22) | -5.69(53.15) | 1.42(45.61) | 6.39(50.30) | 15.49(43.72) | 5.53(44.08) | -0.78(38.78) | 0.81(44.14) | -0.55(52.96) |
| dMONEY | 0.00 **(0.00) | 0.00**(0.00) | 0.00**(0.00) | 0.00**(0.00) | 0.00**(0.00) | 0.00**(0.00) | 0.00**(0.00) | 0.00**(0.00) | 0.00**(0.00) |
| dUNRATE | 0.27(0.15) | -0.03(0.10) | -0.09(0.08) | 0.02 *(0.08) | 0.03 **(0.07) | -0.02(0.07) | -0.05(0.07) | 0.01 **(0.09) | 0.05(0.10) |
| dINFLATION | 9.46 **(3.22) | 5.65 **(2.94) | 5.39 *(2.52) | 7.77 **(2.49) | 5.59 **(2.15) | 5.02 *(2.45) | 3.88(2.49) | 4.47(3.06) | 6.86 *(3.30) |
| dWUPI | 0.19(1.35) | 0.73 **(0.88) | 0.40 *(0.77) | 0.40 *(0.74) | 0.24(0.69) | -0.18(0.77) | -0.44(0.89) | -1.03(1.4) | -0.73 *(0.49) |
| dGPE | -0.01(0.15) | 0.07(0.10) | 0.09(0.08) | 0.05 **(0.08) | 0.13 *(0.07) | 0.09**(0.07) | 0.07**(0.08) | -0.02*(0.1) | 0.01(0.11) |
| dVIX | -25.95 **(7.95) | -5.52(8.24) | -4.77(6.09) | -15.5 **(5.45) | -15.16 **(4.34) | -15.37 **(4.40) | -14.20(4.38) | -16.88 **(4.8) | -1.68(5.71) |
| dGPR | 0.03 (0.02) | 0.02 (0.02) | 0.01 (0.01) | 0.01 (0.01) | 0.01 (0.01) | 0.02 (0.01) | 0.02 (0.01) | 0.02 (0.01) | -0.00(0.01) |
| dSPREAD | -1.58(62.22) | -14.13(48.59) | -6.88(43.13) | -11.89(45.62) | -11.59(41.98) | -12.39(45.34) | 1.94(46.81) | -1.43(64.09) | 1.07(70.29) |
| dGEPU | 0.03**(0.01) | 0.00(0.01) | -0.00(0.01) | -0.01(0.02) | -0.01(0.01) | -0.02*(0.01) | -0.02*(0.01) | -0.04*(0.02) | -0.08**(0.03) |
| **Intercept** | -3.99**(0.95) | -1.84**(0.71) | -0.85**(0.62) | -0.69(0.58) | -0.02(0.05) | 1.44*(0.57) | 2.30**(0.59) | 4.06**(0.78) | 7.20**(0.76) |
| **Pseudo R2** | 0.79 | 0.73 | 0.70 | 0.68 | 0.65 | 0.63 | 0.62 | 0.64 | 0.71 |

Note: Cell entries are coefficients, with standard errors in parentheses, * denotes p < 0:05; ** denotes p < 0:01

Table 10. Parameter estimate of Pandemic and Geo-Political unrest on ROC

| Variable | Qn (0.1) | Qn (0.2) | Qn (0.3) | Qn (0.4) | Qn (0.5) | Qn (0.6) | Qn (0.7) | Qn (0.8) | Qn (0.9) |
|---|---|---|---|---|---|---|---|---|---|
| erSP | 0.66 (0.38) | 0.86 **(0.27) | 0.60 *(0.25) | 0.30(0.24) | 0.14(0.25) | 0.07(0.24) | 0.46 **(0.07) | 0.30(0.28) | -0.18(0.47) |
| dPROD | 0.58(0.44) | -0.15(0.57) | 1.50*(0.65) | 1.67**(0.60) | 0.85(0.27) | 1.69**(0.57) | 0.19 *(0.59) | 0.73 **(0.20) | 0.73 *(0.27) |
| dCURRENCY | -28.85(66.19) | -0.33(92.07) | 0.10(70.72) | 0.10(70.72) | 16.98(43.21) | 14.26(41.79) | -0.52(39.73) | 1.18(49.41) | -0.55(0.72) |
| dMONEY | 0.00 **(0.00) | 0.00 **(0.00) | 0.20 *(0.07) | 0.18 *(0.07) | 0.07 (0.07) | 0.00 **(0.00) | 0.17 *(0.08) | 0.13 (0.10) | 0.26(0.19) |
| dUNRATE | 0.03(0.16) | 0.18*(0.07) | 0.15 *(0.06) | 0.15**(0.05) | -0.01(0.08) | 0.08*(0.03) | 0.09(0.08) | 0.08(0.08) | 0.16*(0.06) |
| dINFLATION | 5.82 **(3.36) | 9.74 *(3.65) | 15.08 **(3.27) | 13.10 **(3.0) | 10.77 **(2.97) | 7.78 *(2.82) | 9.43**(3.36) | 8.80 *(4.36) | 9.93(7.48) |
| dWUPI | 0.83(1.97) | -2.87(2.09) | -4.31**(1.47) | -6.05 **(1.43) | -3.18*(1.18) | -0.66(1.20) | -2.42(1.52) | 0.14(1.87) | 0.55(4.00) |
| dGPE | 37*(19.42) | 69.95**(13.73) | 71.63**(14.21) | 71.03**(11.93) | 65.95**(10.97) | 60.00**(9.41) | 60.66**(9.71) | 60.94**(11.62) | 75.96**(25.87) |
| dVIX | -34.59*(13.53) | -13.33 *(7.14) | -18.69**(6.79) | -27.22 **(6.09) | -16.79 **(5.66) | -23.27 **(5.25) | -21.14 **(4.77) | -22.65 **(7.72) | -25.52 **(11.25) |
| dGPR | 0.13 *(0.05) | 0.11**(0.03) | -0.07*(0.03) | -0.06*(0.03) | 0.01(0.01) | 0.021(0.01) | -0.01(0.01) | -0.00(0.04) | 0.01(0.06) |
| dPSREAD | 6.90(6.55) | -8.33*(3.41) | -0.11(52.62) | -1.94(4.08) | 2.56(3.80) | 6.64 (3.46) | 4.61(50.40) | 8.44(5.97) | 4.97(11.21) |
| dGEPU | -0.00(0.06) | -0.00(0.03) | -0.02(0.02) | -0.02(0.02) | -0.03(0.02) | -0.02(0.02) | -0.02(0.02) | -0.01(0.02) | -0.06(0.03) |
| d(WUPI*GPR) | -0.05*(0.02) | 0.03**(0.01) | 0.06**(0.01) | 0.07**(0.01) | 0.03*(0.01) | 0.00(0.01) | 0.02(0.02) | 0.02(0.01) | 0.01(0.02) |
| **Intercept** | -5.91**(1.76) | -5.14**(0.85) | -3.14**(0.87) | -1.55*(0.82) | 0.05(0.78) | 2.00*(0.75) | 3.25**(0.83) | 3.80**(0.91) | 6.84**(1.47) |
| **Pseudo R2** | 0.76 | 0.67 | 0.64 | 0.61 | 0.59 | 0.57 | 0.58 | 0.62 | 0.69 |

Note: Cell entries are coefficients, with standard errors in parentheses, * denotes p < 0:05; ** denotes p < 0:01

Table 11. Parameter estimate of Geo-Political unrest & Inflation on ROC

| Variable | Qn (0.1) | Qn (0.2) | Qn (0.3) | Qn (0.4) | Qn (0.5) | Qn (0.6) | Qn (0.7) | Qn (0.8) | Qn (0.9) |
|---|---|---|---|---|---|---|---|---|---|
| erSP | 0.33*(0.15) | 0.34*(0.16) | 0.33 *(0.13) | 0.37 **(0.11) | 0.44 **(0.12) | 0.49 **(0.07) | 0.50 **(0.07) | 0.64 **(0.10) | 0.47**(0.12) |
| dPROD | 0.81(0.53) | 0.88(0.81) | 0.92(0.65) | 0.88(0.55) | 1.29*(0.56) | 0.94 **(0.29) | 1.31**(0.44) | 0.89 *(0.47) | 0.98**(0.25) |
| dCURRENCY | 0.33(0.64) | 0.74(0.69) | 1.12*(0.54) | 0.89(0.48) | 0.94*(0.5) | 4.04(43.67) | 0.61(0.38) | -0.51(50.42) | 0.55(0.49) |
| dMONEY | 0.51** (0.08) | 0.22*(0.10) | 0.02 (0.08) | 0.02(0.07) | 0.04(0.07) | 0.00 **(0.00) | 0.06(0.06) | 0.00 **(0.00) | 0.15(0.10) |
| dUNRATE | 0.33*(0.12) | 0.34**(0.09) | 0.39**(0.08) | 0.41**(0.05) | 0.31**(0.04) | -0.04(0.07) | 0.19**(0.03) | -0.00(0.09) | 0.19**(0.03) |
| dINFLATION | 16.99**(4.06) | 17.57**(3.81) | 5.47 *(2.49) | 7.38*(3.09) | 7.84*(3.36) | 4.67 *(2.39) | 7.90**(2.63) | 3.82(3.34) | 9.53*(3.60) |
| dWUPI | 1.28(1.35) | 1.33(1.66) | 1.25(0.96) | 1.01(0.66) | 0.42(1.02) | -0.29(0.74) | -0.30(0.72) | -1.05(1.47) | -0.81(0.49) |
| dGPE | -0.05(0.18) | -0.02(0.14) | 0.08(0.11) | 0.08(0.07) | 0.09(0.10) | 0.07(0.08) | 0.05(0.09) | 0.00(0.12) | 0.11(0.12) |
| dVIX | 0.00(0.01) | 0.01(0.01) | 0.02*(0.00) | 0.02*(0.00) | 0.02*(0.00) | -14.85 **(4.25) | 0.01*(0.00) | -15.76 **(4.66) | 0.02(0.01) |
| dGPR | 0.00(0.05) | -0.05(0.04) | -0.03(0.03) | -0.01(0.03) | -0.01(0.01) | 0.02(0.01) | 0.01(0.01) | 0.02(0.01) | -0.02(0.03) |
| dSPREAD | 8.03(7.81) | 7.48(7.08) | 8.08(5.61) | 3.26(4.93) | 2.69(5.44) | -0.50(4.39) | -0.50(4.39) | -0.57(67.31) | 1.95(6.14) |
| dGEPU | 0.03(0.03) | -0.00(0.03) | -0.03(0.02) | -0.04*(0.07) | 0.00(0.02) | -0.02(0.01) | -0.01(0.02) | -0.02(0.02) | -0.03(0.04) |
| d(GPR*INFLATION) | -0.27*(0.13) | 0.26*(0.11) | -0.08(0.06) | -0.07(0.05) | -0.05(0.08) | -0.03(0.06) | -0.08(0.06) | -0.09(0.06) | -0.02(0.22) |
| **Intercept** | -3.42**(1.07) | -1.87(1.13) | -0.27(0.91) | 0.86(0.57) | 2.41**(0.86) | 3.39**(0.56) | 4.62**(0.71) | 4.82**(0.84) | 11.09**(1.60) |
| **Pseudo R2** | 0.76 | 0.65 | 0.61 | 0.58 | 0.54 | 0.53 | 0.52 | 0.55 | 0.71 |

Note: Cell entries are coefficients, with standard errors in parentheses, * denotes p < 0:05; ** denotes p < 0:01